\documentclass{elsart}

\usepackage{amssymb,amsmath}
\def\equ#1{(\ref{#1})}
\def\cal{\mathcal}
\begin{document}
\begin{frontmatter}
\title{Dual mapping between the 
antisymmetric tensor matter field and the Kalb-Ramond field} 
\author[uece]{L. Gonzaga Filho} 
\author[uece]{M. S. Cunha}
\author[ufc]{R. R. Landim}
   \address[uece]{Grupo de F\'{\i}sica Te\'orica - GFT, Universidade Estadual do Cear\'a -
Av. Paranjana, 1700, CEP 60740-000, Fortaleza, Cear\'a, Brazil}
\address[ufc]{Departamento de F\'{\i}sica, Universidade Federal do
Cear\'a - Caixa Postal 6030, CEP 60455-760, Fortaleza, Cear\'a, Brazil}
\thanks[renan]{renan@fisica.ufc.br}

\begin{abstract}
In this work we present a dual mapping between the Kalb-Ramond and antisymmetric tensor matter (ATM) field actions. Our procedure shows that the correlation functions associated with both the Noether current and the topological current are equivalent.
\end{abstract}
\end{frontmatter}

Kalb-Ramond fields first appeared as a tensorial generalization of vector gauge fields. These allow the construction of topological invariants in $D$-dimensional manifolds \cite{vi.ogievetsky-yf4,m.kalb-prd9,as.schwarz-tca2} and have an important role in dualization \cite{a.smailagic-prd61,a.smailagic-plb489a,e.harikumar-mpla15,r.menezes-plb564}. Such antisymmetric tensor fields are the key to generate mass for the vector gauge fields through the topological mass mechanism \cite{tj.allen-mpla6,i.oda-plb234,rr.landim-plb504}. More recently one finds some applications in the socalled physics of extra dimension. It is also noteworthy that the Kalb-Ramond fields appear in effective theories of specific low energy superstring models, and may describe axion physics or torsion of a Riemannian manifold \cite{sengupta}.

Different types of (second rank) antisymmetric tensor fields have been introduced by Avdeev and Chizhov  in \cite{lv.avdeev-plb321}. There, on the other hand, the action was constructed with a matter field rather than a gauge field exhibiting several interesting features. As showed later by Lemes {\it et al }\cite{v.lemes-plb344}, in a BRST framework, some especific model which have antisymmetric tensor matter (ATM) fields is renormalizable to arbitrary perturbative orders. In a subsequent paper they further realized that the ATM  field is a real component of a complex tensor field that satisfies a complex self-dual condition \cite{v.lemes-plb352}. As shown by the authors this condition makes the ATM field massless. 

In a more recent paper \cite{l.gonzaga-el69} we proposed a mechanism to generate mass to the ATM field which preserves the U(1) symmetry. By this mechanism, a topological term is introduced via a complex vector field. In \cite{l.gonzaga-plb646} we also analyzed the possibility to give mass to a ATM field through the Higgs mechanism. There, a scalar field is coupled to the ATM field and requiring parity conservation it is described as a doublet where one of its components is a pseudo-scalar. We also showed that a topological term for the ATM field can be also generated by spontaneous symmetry breaking.

In spite of a large variety of papers thereafter published about these two types of tensors, no one has reported a possible connection between them. In the present paper we will show the existence of a dual mapping between the Kalb-Ramond and ATM fields.  We apply the dual mapping method developed by Fosco \textit{et al.} \cite{cd.fosco-ap290} by making use of the complex self-dual condition and the $U(1)$ symmetry.

Let us consider the action  for the ATM field  written in terms of the complex self-dual tensor \cite{v.lemes-plb352}

\begin{equation}
S(\varphi,\varphi^{\dagger})= \int d^4x\, \partial_{\mu}\varphi^{\mu\nu}\partial^{\rho}\varphi_{\rho\nu}^{\dagger}, 
\label{ST}
\end{equation}  
where $\varphi_{\mu\nu}=T_{\mu\nu} +i\widetilde{T}_{\mu\nu}$ is the complex anti-symmetric tensor field satisfying the complex self-dual condition  $\varphi_{\mu\nu}=i\widetilde{\varphi}_{\mu\nu}$ \cite{v.lemes-plb352}. In order to show the dual mapping procedure we first linearize the derivatives of the kinetic term with the introduction of a complex field $b_\mu=c_\mu+id_\mu$. We can now write the action \equ{ST} in the form
     
\begin{equation}
S(\varphi,b)= \int d^4x\left(b_{\mu}^{\dagger}\partial_{\rho}\varphi^{\rho\mu} + b^{\mu}\partial_{\rho}\varphi^{\rho\mu\dagger} - b^{\mu}b_{\mu}^{\dagger} \right).
\label{Sb}
\end{equation}
The on-shell equivalence of  (\ref{ST}) and (\ref{Sb}) can be obtained by eliminating  $b_{\mu}$  through the equation of motion 
\begin{equation}
\frac{\delta S}{\delta b^{\dagger\mu}}= \partial^{\rho}\varphi_{\rho\mu} - b_{\mu}=0.
\end{equation}

The Noether current related to the global $U(1)$ invariance of Eq. \equ{Sb} is obtained as usual by replacing the normal derivative of the tensor field $\varphi_{\mu\nu}$ by its covariant derivative  $D_\mu\varphi_{\rho\sigma}=\partial_{\mu}\varphi_{\rho\sigma}+is_\mu\varphi_{\rho\sigma}$, namely
\begin{equation}
J^{\mu}=i\left(b_{\nu}^{\dagger}\varphi^{\mu\nu} - b_{\nu}\varphi^{\mu\nu\dagger}\right).
\end{equation}
where $s_\mu$ is an external source.

The generating  functional $\mathcal{Z}[s_\mu]$  for current-current correlation is
\begin{equation}
{\cal Z}[s_{\mu}]=\int [D\varphi Db] e^{-i\left( S(\varphi,b)\;+\;\int d^4x\,s_{\mu}J^{\mu}\right)},\label{Z}
\end{equation}
where the invariant functional measure can be written as
\begin{equation}
 [D\varphi Db]=\delta(\varphi_{\mu\nu}-i\widetilde{\varphi}_{\mu\nu}) D\varphi_{\mu\nu}D\varphi_{\mu\nu}^{\dagger}Db_{\mu}Db_{\mu}^{\dagger}.
\end{equation}
From gauge invariance of \equ{Z}, it follows that
\begin{equation}
{\cal Z}[s_\mu]=\int [D\varphi Db] D\eta_{\mu}\delta[{f}_{\mu\nu}(\eta)] e^{-i\left( S(\varphi,b)\;+\;\int d^4x(s_{\mu}+\eta_\mu)J^{\mu}\right)}.
\label{invZ}
\end{equation}
Here we have defined ${f}_{\mu\nu}=\varepsilon_{\mu\nu\rho\sigma}\partial^{\rho}\eta^{\sigma}$, $\eta_{\mu}=\partial_{\mu}\alpha$, and $\alpha$ is a scalar field.

With the following representation of the Dirac's delta functional
\begin{equation}
\delta[{f}_{\mu\nu}(\eta)]=\int DB_{\mu\nu} e^{-i \int d^4x B_{\mu\nu}\varepsilon^{\mu\nu\rho\sigma}\partial_{\rho}\eta_{\sigma}}\;,  
\end{equation}
where $B_{\mu\nu}$ is an anti-symmetric tensor field, Eq. \equ{invZ} can be written as
\begin{eqnarray}
Z[s_{\mu}] \!=\! \int[D\varphi Db] DB_{\mu\nu}D\eta_{\mu}e^{-i\left( S(\varphi,b)\;+\;\int d^4x(s_{\mu}+\eta_\mu)J^{\mu}+\int d^4x B_{\mu\nu}\varepsilon^{\mu\nu\rho\sigma}\partial_{\rho}\eta_{\sigma}\right)}  . 
\end{eqnarray}  
Redefining $\eta_\mu$ by $\eta_\mu=v_\mu-s_\mu$, the external source $s_{\mu}$ decouples from the matter field, and we get

\begin{eqnarray}
Z[s_{\mu}] = \int[D\varphi Db] DB_{\mu\nu}Dv_{\mu}e^{-i\left( S(\varphi,b)\;+\;\int d^4x v_{\mu}J^{\mu}+ \int d^4x B_{\mu\nu}\varepsilon^{\mu\nu\rho\sigma}\partial_{\rho}(v_{\sigma}-s_\sigma)\right)}  
\end{eqnarray}
We can now define a dual action for $B_{\mu\nu}$ as follows

\begin{equation}
e^{-iS_{\rm dual}[B_{\mu\nu}]} = \int[D\varphi Db]Dv_{\mu}e^{-i\left( S(\varphi,b)\;+\;\int d^4x v_{\mu}J^{\mu}+ \int d^4x B_{\mu\nu}\varepsilon^{\mu\nu\rho\sigma}\partial_{\rho}v_{\sigma}\right)},
\label{ZB}
\end{equation}
such that
\begin{equation}
{\cal Z}[s_\mu]=\int DB_{\mu\nu} e^{-i\left(S_{\rm dual}[B_{\mu\nu}]- \int d^4x s_{\mu}
\epsilon^{\mu\nu\rho\sigma}\partial_{\nu}B_{\rho\sigma}\right) \;}.
\label{Zdual}
\end{equation}
As we can see from Eq. \equ{ZB}, $B_{\mu\nu}$ is in fact a true gauge field with a gauge invariant action
\begin{equation}                                                                        
S_{{\rm dual}}[B_{\mu\nu}]\;=\;S_{\rm dual}[B_{\mu\nu} + \partial_{\mu}\omega_{\nu}]\;.
\label{gin}
\end{equation}    
The $U(1)$ global symmetry for the fields and the complex self-dual condition are the key ingredients to implement our proposed duality. Actually, it is a well-known fact that the existence of a global symmetry is crucial to obtain dualities. 

Let us underline here that the same generating functional given by the equations \equ{Z} and \equ{Zdual} written in terms of $\varphi_{\mu\nu}$ or $B_{\mu\nu}$ gives the same correlation function for the $U(1)$ current and the topological current, \textit{i.e.},

\begin{equation}
\left\langle J_{\mu _1}(x_1)J_{\mu
    _2}(x_2)\;\ldots\;J_{\mu _n}(x_n)\right\rangle _{\varphi_{\mu\nu}} \label{corr1}
=\;\left\langle j_{\mu _1}^T(x_1)j_{\mu _2}^T(x_2)\;\ldots\;
j_{\mu_n}^T(x_n)\right\rangle _{B_{\mu\nu}}\; ,
\end{equation}
where
\begin{equation}
\label{mapdual1}
 j_\mu^T=\epsilon^{\mu\nu\rho\sigma}\partial_{\nu}B_{\rho\sigma},
\end{equation}
and

\begin{equation}
J_{\mu}=i\left(\partial_{\rho}\varphi{\rho\nu}^{\dagger}\varphi_{\mu\nu} - \partial_{\rho}\varphi^{\rho\mu}\varphi_{\mu\nu\dagger}\right)= 2(\partial_{\rho}\widetilde{T}^{\rho\nu}T_{\mu\nu} - \partial_{\rho}T^{\rho\nu}\widetilde{T}_{\mu\nu}).
\end{equation}
Note that both $j_\mu^T$ and $J_{\mu}$ are axial currents. As a consequence, parity is preserved in this duality.  It is worth mentioning that the dual mapping developed in \cite{cd.fosco-ap290} breaks the parity symmetry for that scalar model in three dimension.

In conclusion, in this paper we have obtained a dual mapping between the Kalb-Ramond  and the ATM field actions. This kind of dualization was implemented by the requirement of complex self-duality. As we saw, the same correlation function for the $U(1)$ current and the topological currentthe was obtained. As a consequence parity is preserved and it is free of axial anomalies.

\section*{Acknowledgment}
We wish to thank H.R Christiansen for a critical reading of the manuscript. The Conselho Nacional de Desenvolvimento Cient\'\i fico e tecnol\'ogico-CNPq is gratefully acknowledged for financial support.

\section*{Dedicatory} R. R. Landim - This paper is dedicated to the memory of my wife, Isabel Mara.

\end{document}